\documentclass[5p]{elsarticle}

\usepackage{amssymb}
\usepackage{amsmath}
\usepackage{graphicx}
\usepackage{latexsym}
\usepackage{amsfonts}
\usepackage{color,dsfont}

\begin{document}

\title{$\mathcal{PT}$-symmetric ladders with a scattering core}
\author{J. D'Ambroise}
\address{Department of Mathematics, Amherst College, Amherst, MA 01002-5000, USA}
\author{S. Lepri}
\address{CNR-Consiglio Nazionale delle Ricerche, Istituto dei Sistemi Complessi, via
Madonna del piano 10, I-50019 Sesto Fiorentino, Italy}
\address{Istituto Nazionale di Fisica Nucleare, Sezione di Firenze, 
via G. Sansone 1, I-50019 Sesto Fiorentino, Italy}
\author{B. A. Malomed}
\address{Department of Physical Electronics, School of Electrical Engineering,
Faculty of Engineering, Tel Aviv University, Tel Aviv 69978, Israel}
\author{P.G. Kevrekidis}
\address{Department of Mathematics and Statistics, University of Massachusetts,
Amherst, MA 01003-9305, USA}

\begin{abstract}
We consider a $\mathcal{PT}$-symmetric chain (ladder-shaped) system governed by the discrete nonlinear Schr\"{o}dinger equation where the cubic nonlinearity is carried solely by two central ``rungs" of the ladder. Two branches of scattering solutions for incident plane waves are found. We systematically construct these solutions, analyze their stability, and discuss non-reciprocity of the transmission associated with them. To relate the results to finite-size wavepacket dynamics, we also perform direct simulations of the evolution of the wavepackets, which confirm that the transmission is indeed asymmetric in this nonlinear system with the mutually balanced gain and loss.
\end{abstract}

\begin{keyword}
Discrete nonlinear Schr\"odinger equation, $\mathcal{PT}$-symmetry, cubic nonlinearity
\PACS {05.45.-a, 63.20.Ry}
\end{keyword}

\maketitle


\section{Introduction}

A powerful tool for the control of the energy transfer in chain-like systems
is provided by settings which are capable to induce asymmetric
(nonreciprocal) wave propagation in such systems, i.e., \textit{wave diodes}
. In particular, the asymmetric phonon transmission through a nonlinear
interface between dissimilar crystals was reported in Ref. \cite
{Kosevich1995}. Acoustic-wave diodes have been demonstrated in nonlinear
phononic media too \cite{Liang09,Liang2010}. The propagation of acoustic
waves through granular crystals may also be promising in this respect. In
particular, experiments have produced a change of the reflectivity of
solitary waves from the boundary between different granular media \cite
{Nesterenko05}. A related effect of the rectification of the energy transfer
at particular frequencies in a chain of particles with an embedded defect
has been reported in Ref. \cite{Boechler2011}. In nonlinear optics, the
``all-optical diode" was theoretically elaborated in Refs.~
\cite{Scalora94,Tocci95}, which was followed by its experimental realization
\cite{Gallo01}. Other realizations of the unidirectional transmission have
been considered in metamaterials \cite{Feise05}, regular \cite{Konotop02}
and quasiperiodic \cite{Biancalana08} photonic crystals, chains of nonlinear
cavities \cite{Grigoriev2011}, and, quite recently, $\mathcal{PT}$-symmetric
waveguides \cite{Ramezani2010,D'Ambroise2012,Bender2013}. In Ref. \cite
{Roy2010}, an extension for quantum settings, in which the diode effect is
realized in few-photon states, was proposed. It is also relevant to mention
a related work for electric transmission lines \cite{Tao2011}.

A basic model for the implementation of this class of phenomena is a
particular form of the discrete nonlinear Schr\"{o}dinger (DNLS) equations
\cite{Eilbeck1985,Kevrekidis}, in which a finite-size nonlinear core is
embedded into a linear chain \cite
{Brazhnyi2011,Molina1993,Gupta1997,Gupta1997a,Bulgakov2011}. The use of
DNLS-based models is particularly relevant in the present context, as these
models, with a short nonlinear segment inserted into the bulk linear
lattice, make it possible to solve the stationary scattering problem exactly
\cite{Lepri2011}. It has been found that the embedded nonlinearity can be
employed to design a chain operating as a diode, which transmits waves with
the equal amplitudes and frequencies asymmetrically in the opposite
directions \cite{Lepri2011,Lepri2014}. This model can be extended to study
the effect of magnetic flux on the rectification \cite{Marchesoni2014}.

Obviously, the propagation direction favored by a diode chain is reversed in
a mirror-image version of the given system, which suggests to consider the
transmission of waves in dual systems, built of two such parallel chains
with opposite orientations, linearly coupled to each other in the transverse
direction. This ``diode-antidiode'' system was introduced and analyzed in
Ref. \cite{Lepri2013}. It was demonstrated that the increase of the
nonlinearity strength leads to the spontaneous symmetry breaking between the
diode and antidiode cores, thus allowing the transmission of large-amplitude
waves in either direction.

As mentioned above, $\mathcal{PT}$-symmetry chains, which are built of
separated elements carrying equal amounts of linear amplification and
dissipation, also allow one to implement the unidirectional or asymmetric
propagation of waves \cite{Ramezani2010,D'Ambroise2012}. This fact suggests
to introduce the $\mathcal{PT}$-symmetric version of the diode-antidiode
system, and consider the wave transmission in such settings, which is the
subject of the present work. In addition to the specific interest concerning
the relation between the bi- and unidirectional propagation, this system is
a relevant addition to a variety of $\mathcal{PT}$-symmetric discrete
lattices, which have been introduced in recent works \cite{discr-PT},
chiefly in the form of nonlinear discrete dynamical equations. The study of
the existence, stability and dynamical properties of these systems is an
interesting problem in its own right.

The rest of the paper is organized as follows. The model is
formulated in Section II. Stationary solutions of the scattering
problem for asymptotically linear waves impinging on the central
nonlinear core of the system are reported in Section III.A, and their stability is analyzed in Section III.B. With this
nonlinearity, one of the key tasks is to actually construct
standing-wave states in the system, which is done in Section III.
Then, measuring the transmissivity in either direction, we identify
and quantify the transmission asymmetry. Given the extended nature
of these constructed stationary states, in Section IV we address a
more (numerically) quantifiable manifestation of the
nonlinearity-induced asymmetry, simulating the scattering of
finite-size Gaussian wavepackets on the central nonlinear core of
the chain, for either incidence direction. The paper is concluded by
Section V, which also outlines directions for the extension of the
research.

\section{The model}

\label{sec:mod}

\begin{figure}[tbp]
\begin{center}
\includegraphics[width=8cm,angle=0,clip]{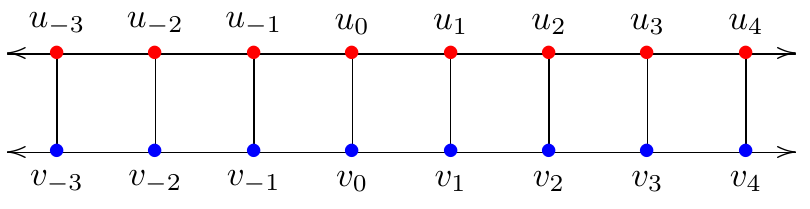}
\end{center}
\caption{The picture shows the ladder configuration.  Red dots on the top side of the ladder correspond to wave function $u_n$ and to a linear gain potential $+i\gamma$.  Blue dots on the bottom side of the ladder correspond to the wave function $v_n$ and to a linear loss potential $-i\gamma$. Vertical lines denote coupling with coefficient $\kappa$ across the rungs of the ladder.  Note that the real parts of the linear potential, $U_n$ and $V_n$, are nonzero only at $n=1,2$.}
\label{fig0}
\end{figure}

Following Ref. \cite{Lepri2011}, which had revealed the possibility of the
asymmetric transmission in nonlinear chains, a ladder-type model of two
linearly coupled chains with opposite directions of transmission was
introduced in Ref.~\cite{Lepri2013}, while the asymmetric transmission in a
linear system with $\mathcal{PT}$-symmetric embedded defects was introduced
in Ref. \cite{D'Ambroise2012}; a different example featuring unidirectional
propagation in a $\mathcal{PT}$-symmetric chain was given in~\cite%
{Ramezani2010}. This fact, as well as the general current interest to the
dynamics of nonlinear $\mathcal{PT}$-symmetric systems, including discrete
ones \cite{discr-PT}, suggests to consider a $\mathcal{PT}$-symmetric
extension of the two-chain model introduced in Ref. \cite{Lepri2013}. The
simplest variant of the system can be adopted in the following form:
\begin{gather}
i\frac{du_{n}}{dz}=-u_{n+1}-u_{n-1}+\kappa v_{n}+i\gamma u_{n}  \notag \\
+U_{n}u_{n}+\lambda \left( \delta _{n,1}+\delta _{n,2}\right)
|u_{n}|^{2}u_{n},  \notag \\
\label{eqns} \\
i\frac{dv_{n}}{dz}=-v_{n+1}-v_{n-1}+\kappa u_{n}-i\gamma v_{n}  \notag \\
+V_{n}v_{n}+\lambda \left( \delta _{n,1}+\delta _{n,2}\right)
|v_{n}|^{2}v_{n},  \notag
\end{gather}%
where the evolutional variable, $z$, is the propagation distance in terms of
the underlying optical model, $\kappa $ is the coefficient of the transverse
linear coupling, and $\gamma $ is the gain-loss coefficient accounting for
the $\mathcal{PT}$ symmetry of the system. Figure \ref{fig0} shows the ladder configuration.  It is assumed that the chains are
uniform and linear, except for the asymmetric (for $\varepsilon \neq 0$)
localized linear potential,
\begin{eqnarray}
&&U_{n}=V_{0}\left[ (1+\varepsilon )\delta _{n,1}+(1-\varepsilon )\delta
_{n,2}\right] ,  \notag \\
&&V_{n}=V_{0}\left[ (1-\varepsilon )\delta _{n,1}+(1+\varepsilon )\delta
_{n,2}\right] ,  \label{Pot}
\end{eqnarray}%
with amplitude $V_{0}$ and the left-right skew-symmetry coefficient, $%
\varepsilon >0$, and the localized self-defocusing onsite nonlinearity with
strength $\lambda ${$>0$}. Note that the compatibility of potential (\ref%
{Pot}) with the $\mathcal{PT}$ symmetry of Eqs. (\ref{eqns}) is obvious for $%
\varepsilon =0$. At $\varepsilon \neq 0$, this depends on the definition of
the $\mathcal{P}$ transformation: the system remains $\mathcal{PT}$%
-symmetric if the full spatial reversal is understood as the combination of
the switch between the parallel chains, $u_{n}\rightleftarrows v_{n}$ (i.e.,
the $\mathcal{P}$ transformation in the vertical direction) and the
reflection in the horizontal direction, with respect to the midpoint between
$n=1$ and $n=2$.

It is also relevant to mention that the uniform linear coupling
(with coefficient $\kappa $) in Eqs. (\ref{eqns}) between the
parallel chains
corresponds to the ``ladder" system, in terms of Ref. \cite%
{Lepri2013}. The other system considered in that work, of the
``plaquette" type, in which the linear coupling was also localized
[cf. Eq. (\ref{Pot})], with $\kappa _{n}=\kappa \,(\delta
_{n,1}+\delta _{n,2})$, is irrelevant in the present setting, as the $%
\mathcal{PT}$ symmetry may only be maintained by $\kappa >\gamma $, see Eq. (%
\ref{omega}) below.

In the linear parts of the system (at $n\neq 1,2$), a solution to Eqs. (\ref%
{eqns}) can be looked for as
\begin{equation}
\left(
\begin{array}{c}
u_{n} \\
v_{n}%
\end{array}%
\right) =\left(
\begin{array}{c}
A \\
B%
\end{array}%
\right) e^{iqz+iKn},  \label{mode}
\end{equation}%
which yields the dispersion relation for the linear waves,
\begin{equation}
q(K)=2\cos K\pm \sqrt{\kappa ^{2}-\gamma ^{2}},  \label{omega}
\end{equation}%
hence the model makes sense as the one supporting the transmission of waves
once the condition of ${0\leq }\gamma <\kappa $ is imposed (i.e., the
gain-loss coefficient should not be too large in comparison with inter-chain
coupling $\kappa $). If the dispersion relation (\ref{omega}) holds, the
relation between the amplitudes in solution (\ref{mode}) is%
\begin{equation}
\kappa B=\left( -i\gamma \mp \sqrt{\kappa ^{2}-\gamma ^{2}}\right) A,
\label{eq: BA}
\end{equation}%
where $\mp $ corresponds to $\pm $ in Eq. (\ref{omega}), the total intensity
of the wave being%
\begin{equation}
\sqrt{\left\vert A\right\vert ^{2}+\left\vert B\right\vert ^{2}}={\sqrt{2}%
\left\vert A\right\vert }.  \label{I}
\end{equation}

\section{Stationary solutions}

\subsection{Plane waves}

Substituting $\{u_{n},v_{n}\}\equiv \{e^{iqz}{}\phi _{n},e^{iqz}\psi
_{n}\}$ in Eq. (\ref{eqns}) gives rise to the full system of
stationary equations:
\begin{equation}
 \label{eq:stat}
\phi _{n-1} = (i\gamma -q+U_{n}+\alpha _{n}|\phi _{n}|^{2})\phi _{n}+\kappa
\psi _{n}-\phi _{n+1},
\end{equation}
\begin{equation}
\psi _{n-1} = (-i\gamma -q+V_{n}+\beta _{n}|\phi _{n}|^{2})\psi
_{n}+\kappa \phi _{n}-\psi _{n+1}. \notag
\end{equation}
We begin our analysis of the system by constructing stationary solutions in
terms of two wave numbers, $K_{1}$\ and $K_{2}$, which correspond,
respectively, to the upper and lower signs in (\ref{omega}).

For $K_{1},K_{2}>0$, i.e., the incident waves arriving from the left, we
look for solutions to Eqs. (\ref{eq:stat}) as 
$\left(
\begin{array}{c}
\phi _{n} \\
\psi _{n}
\end{array}
\right) =$
\begin{equation}
\left\{
\begin{array}{ll}
\left(
\begin{array}{c}
R_{0,u} \\
R_{0,v}%
\end{array}%
\right) e^{iK_{1}n}+\left(
\begin{array}{c}
R_{u} \\
R_{v}%
\end{array}%
\right) e^{-iK_{1}n} & \quad \\
\quad +\left(
\begin{array}{c}
S_{0,u} \\
S_{0,v}%
\end{array}%
\right) e^{iK_{2}n}+\left(
\begin{array}{c}
S_{u} \\
S_{v}%
\end{array}%
\right) e^{-iK_{2}n}, & n\leq 1 \\
\left(
\begin{array}{c}
T_{1,u} \\
T_{1,v}%
\end{array}%
\right) e^{iK_{1}n}+\left(
\begin{array}{c}
T_{2,u} \\
T_{2,v}%
\end{array}%
\right) e^{iK_{2}n}, & n\geq 2%
\end{array}%
\right. .  \label{ansatz}
\end{equation}%
Here $R_{0,\ast },R_{\ast },T_{1,\ast }$ for $\ast =u,v$, are
amplitudes of the the incident, reflected and transmitted waves associated
with the $K_{1}$ wave in the $u$ and $v$ chains, and $S_{0,\ast },S_{\ast
},T_{2,\ast }$ are similar amplitudes associated to the $K_{2}$ wave. For $%
K_{1},K_{2}<0$, one may obtain a mirror-image solution, with potentials $%
U_{n},V_{n}$ flipped across the midpoint between the $n=1,2$ sites [the
latter transformation is mentioned above in the connection to the definition
of the $\mathcal{PT}$ symmetry of Eq. (\ref{eqns}) with $\varepsilon \neq 0$%
]. In this way, ansatz (\ref{ansatz}) applies as well to the negative
wavenumbers.

\begin{figure}[tbp]
\begin{center}
\includegraphics[width=8cm,angle=0,clip]{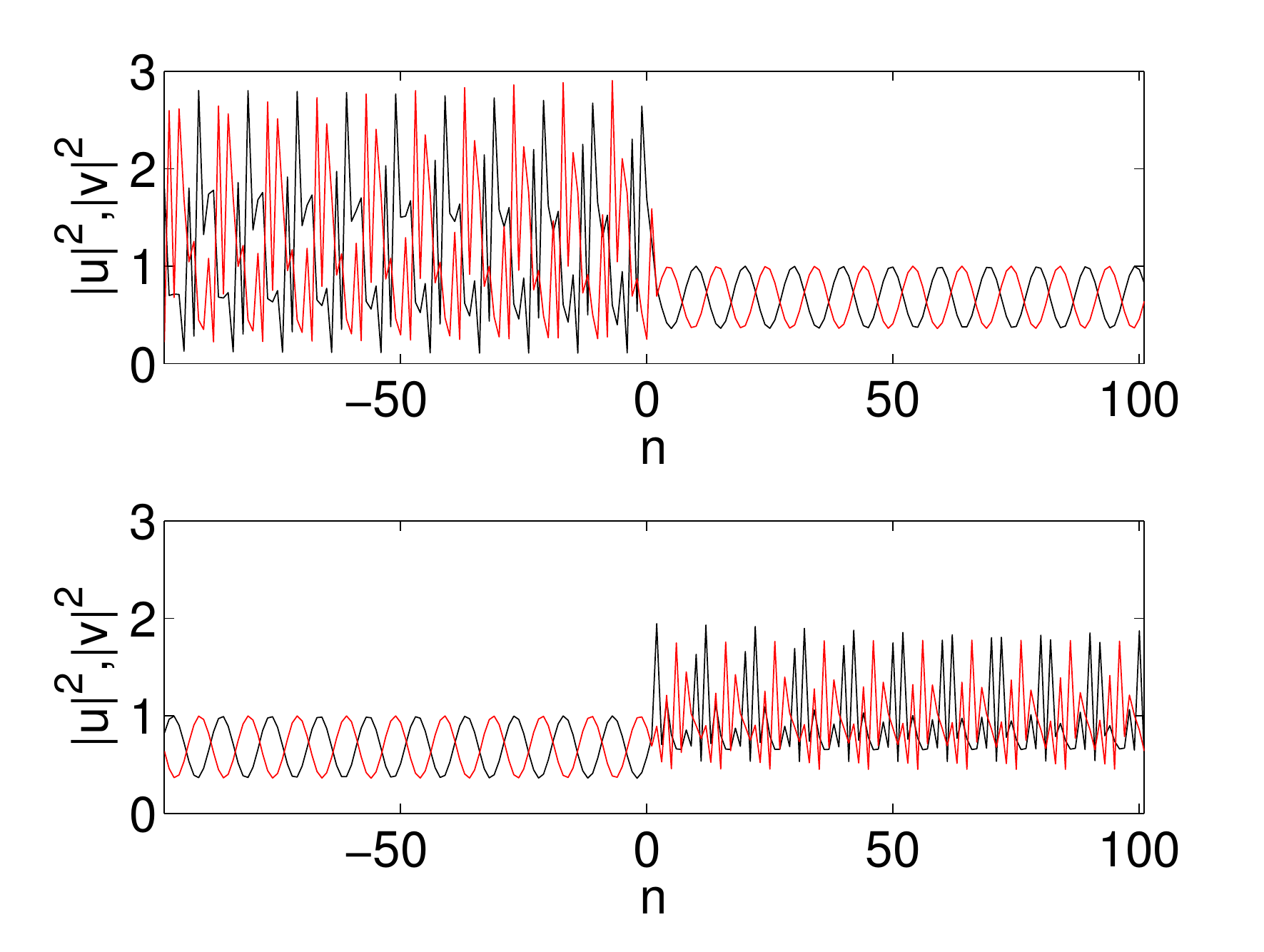}
\end{center}
\par
\caption{(Color online) Stationary-solution profiles of $|u|^{2}, |v|^{2}$ (black, red) 
for $\protect\gamma =0.1$, $\protect\kappa =0.6$, $\protect\varepsilon =0.25$, $%
V_{0}=1$, $T_{1,u}=0.2$, $T_{2,u}=0.8$, $K_{2}=\protect\pi /2$ (top
row), $K_{2}=-\protect\pi /2$ (bottom row), in the linear system ($%
\protect\lambda =0$) with the total ladder length of $200$ for each of $u,v$%
. Amplitudes $R_{0,u},R_{u},S_{0,u},S_{u}$ are obtained by solving the
linear system (\protect\ref{eq:stat}) for $n=1,2$. } \label{profLIN}
\end{figure}

\begin{figure}[tbp]
\begin{center}
\includegraphics[width=8cm,angle=0,clip]{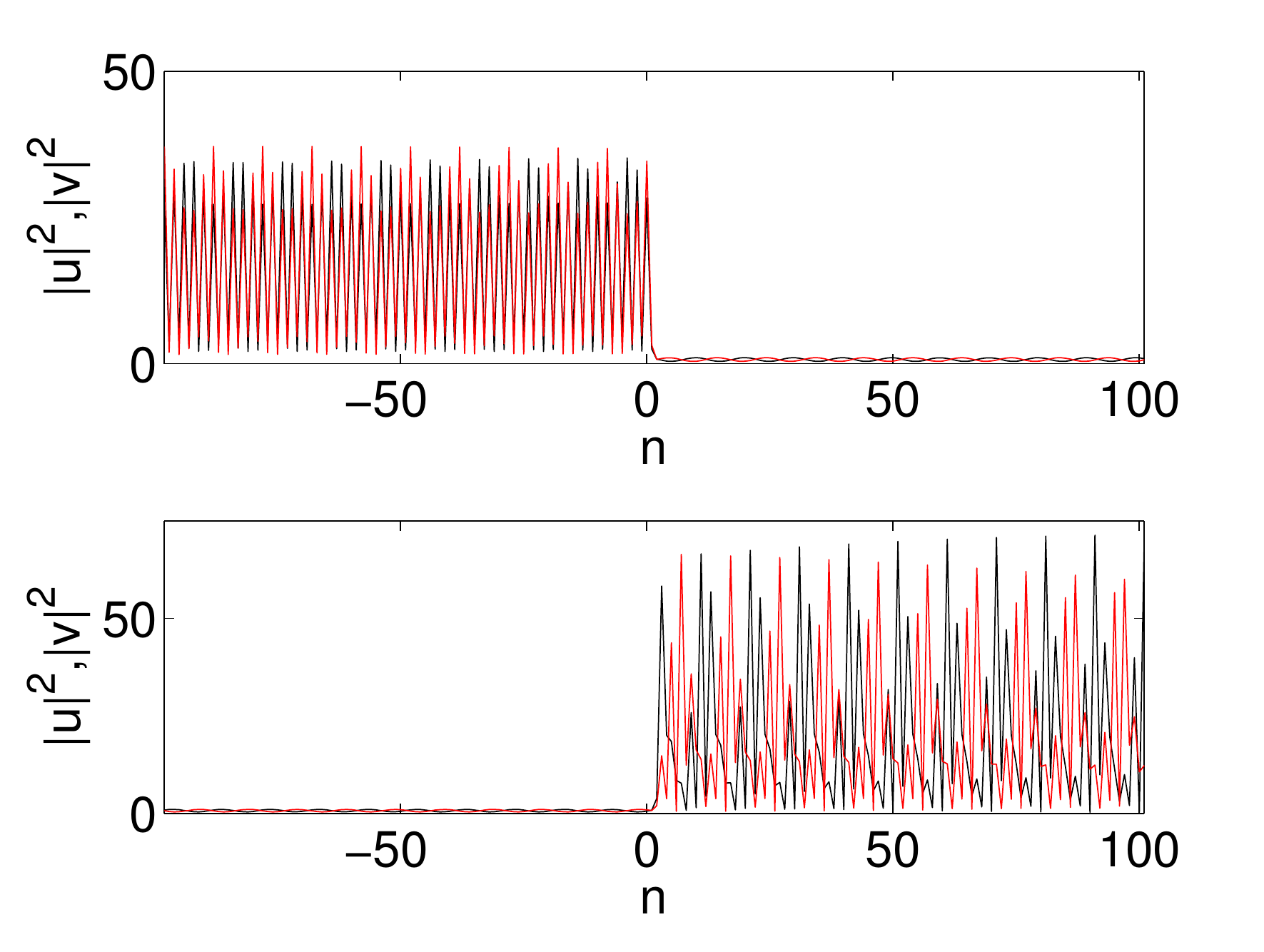}
\end{center}
\caption{(Color online) The nonlinear version of Fig. \protect\ref{profLIN},
with $\protect\lambda =1$. }
\label{profNLIN}
\end{figure}

\begin{figure}[tbp]
\centerline{
\includegraphics[width=7.5cm,angle=0,clip]{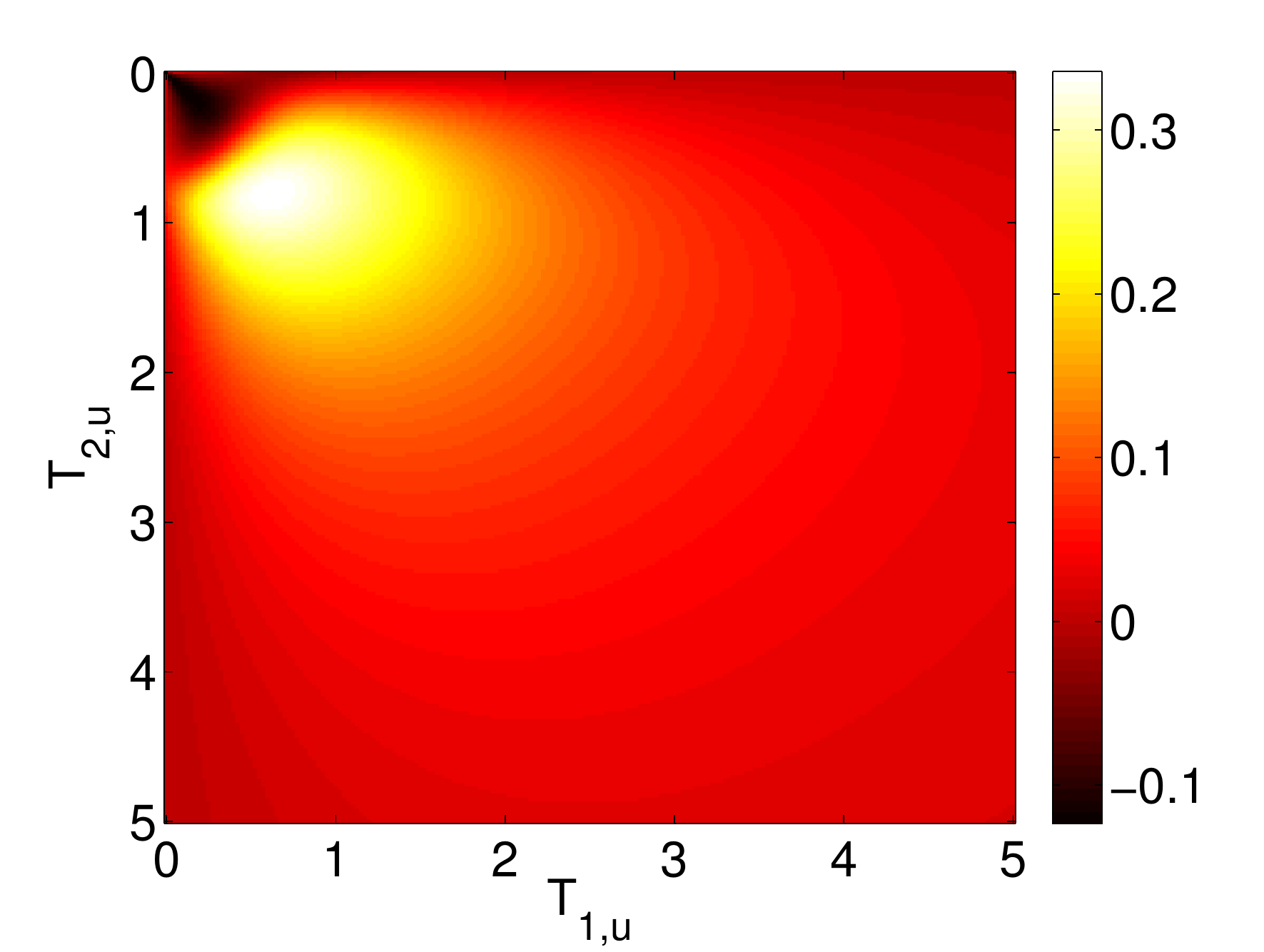}
}
\caption{(Color online) The rectification factor $f$ in equation (\protect
\ref{rectf}) with $t$ defined as per Eq. (\protect\ref{t_amp}), its nonzero
values indicating the non-reciprocity. If $t$ is computed, instead, as per
Eq. (\protect\ref{t_norm}), the plot of $f$ is visually indistinguishable
from the above plot. The parameters are $\protect\lambda =1$, $\protect%
\gamma =0.1$, $\protect\kappa =0.6$, $\protect\varepsilon =0.25$, $K_1 = \pi/2$, $V_{0}=1$, $%
\Delta T_{1,u}=\Delta T_{2,u}=0.025$. }
\label{fNLIN}
\end{figure}

\begin{figure}[tbp]
\begin{center}
\includegraphics[width=8cm,angle=0,clip]{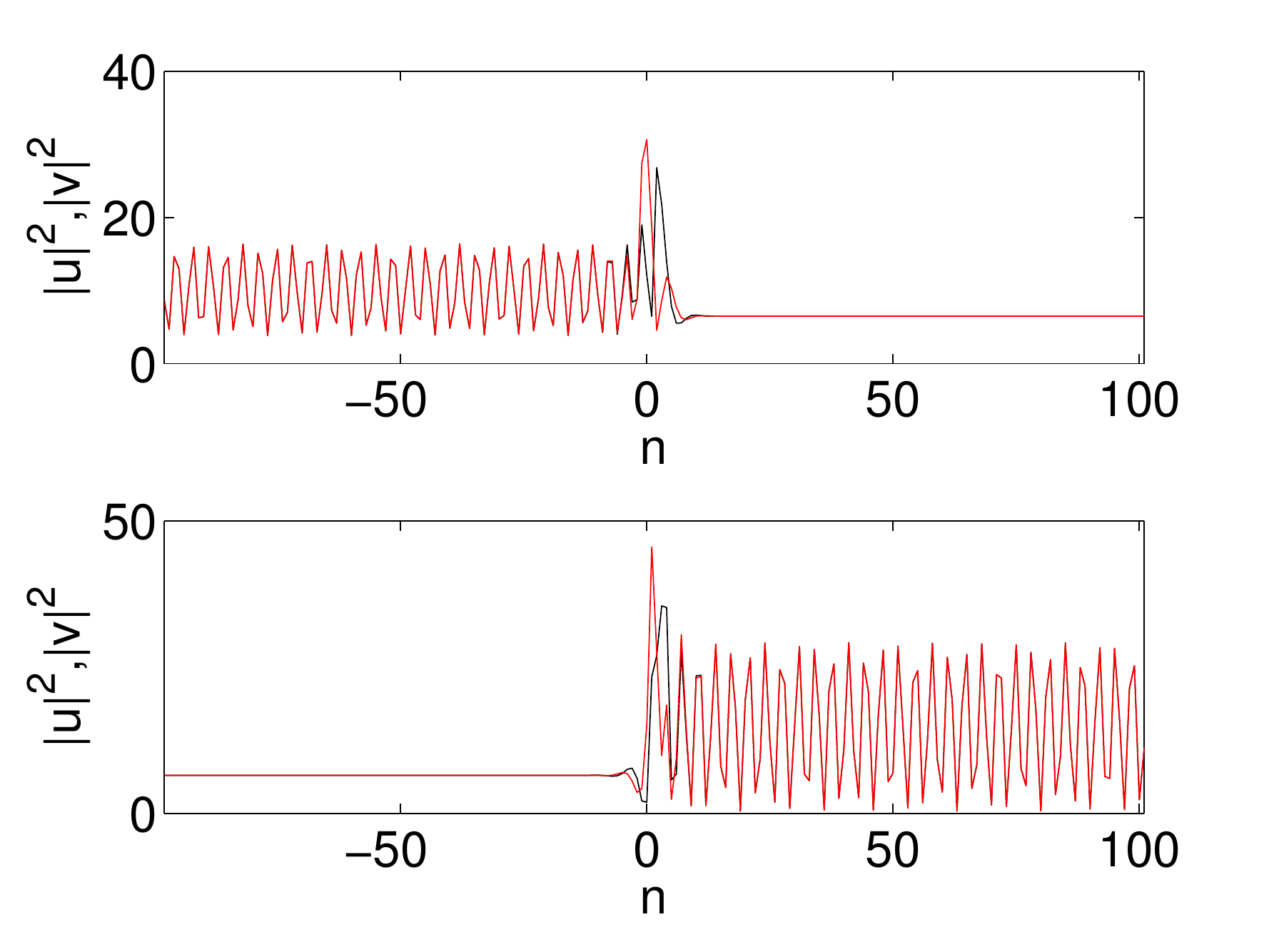}
\end{center}
\caption{(Color online) Stationary-solution profiles of $|u|^{2},|v|^{2}$ (black, red)
with $K_{2}$ real and $K_{1}$ complex. The parameters are $q=-1.8$, $\protect%
\gamma =0.8$, $\protect\kappa =1$, $\protect\varepsilon =0.2$, $V_{0}=1$, $%
R_{0,u}=0$, $S_{0,u}=3$, and $\protect\lambda =0$. Amplitudes $T_{1,u}$, $%
R_{u},T_{2,u}$, $S_{u}$ are obtained by solving linear system (\protect\ref%
{eq:stat}) for $n=1,2$.  For the top plot $K_{1}\approx \protect\pi
+0.62i,K_{2}\approx 2.21$ are determined by Eq.
(\protect\ref{omega}) and for the bottom plot $K_{1}\approx -(\pi
+0.62i),K_{2}\approx -2.21$. Here the lattice length is $200$.
\label{profLINKcomplex}
}
\end{figure}

In fact, one can introduce the initial ansatz so that Eq.
(\ref{ansatz}) holds only for the sites $n<1$ and $n>2$, i.e., in
the linear parts of the system. Then, plugging the so restricted
ansatz into Eq. (\ref{eq:stat}) shows that the full expression
(\ref{ansatz}) follows as a consequence. In other words, the format
of the ansatz applies at $n=1$ and $n=2$ if one originally defines
it solely at $n<1$ and $n>2$.

For each of the summands in Eq. (\ref{ansatz}), the relation between the
amplitudes in the form of Eq. (\ref{eq: BA}) applies. Namely, amplitude
pairs $\{A,B\}=\{R_{0,u},R_{0,v}\}$, $\{R_{u},R_{v}\}$, $\{T_{1,u},T_{1,v}\}$ with
$K_{1}$ satisfy Eq. (\ref{eq: BA}) with the upper (minus) sign, while
amplitude pairs $\{A,B\}=\{S_{0,u},S_{0,v}\}$, $\{S_{u},S_{v}\}$, %
$\{T_{2,u},T_{2,v}\}$ with $K_{2}$ satisfy Eq. (\ref{eq: BA}) with the lower
(plus) sign. In other words, all amplitudes of $\psi $ can be computed in
terms of the amplitudes of $\phi $. We consider $q,\kappa ,\gamma $ as
control parameters, $K_{1},K_{2}$ being computed via Eq. (\ref{omega}). To
determine the amplitudes in the stationary solution in the form of Eq. (\ref%
{ansatz}), one begins by eliminating all the $\psi $ amplitudes in
favor of their $\phi $ counterpart, as per the above relations.
Then, specifying values for two out of the six $\phi $ amplitudes,
we solve the equations for the remaining four $\phi $ amplitudes
using four equations (\ref{eq:stat}) at $n=1,2$. In the linear case
($\lambda =0$), this yields four complex linear equations. In the
nonlinear case, with $\lambda \neq 0$, the equations are linear only
if $|\phi _{1}|^{2},|\psi _{1}|^{2},|\phi _{2}|^{2},|\psi _{2}|^{2}$
are known. In other words, in the nonlinear case we first compute
the input as a function of the transmitted output by specifying the
two $\phi $ amplitudes, $T_{1,u}$ and $T_{2,u}$, and then
solve for the remaining four $\phi $ amplitudes $R_{0,u},R_{u},S_{0,u},S_{u}$%
. In the linear case, one may either compute the input as a function
of the output, like in the nonlinear case, or first specify the
input, $R_{0,u}$ and $S_{0,u}$, and subsequently use Eq.
(\ref{eq:stat}) at $n=1,2$ to solve for the output, \textit{viz}.,
$R_{u},S_{u},T_{1,u},T_{2,u}$. The so obtained sample profiles in
the linear and nonlinear cases are shown in Figs. \ref{profLIN} and
\ref{profNLIN}.

For the solution in the form of Eq. (\ref{ansatz}), the transmission
coefficient can be defined either locally,
\begin{equation}
t=\frac{|T_{1,u}|^{2}+|T_{2,u}|^{2}}{|R_{0,u}|^{2}+|S_{0,u}|^{2}},
\label{t_amp}
\end{equation}
or globally,
\begin{equation}
t=\frac{\displaystyle\sum_{n\geq
3}|T_{1,u}e^{iK_{1}n}+T_{2,u}e^{iK_{2}n}|^{2}}{\displaystyle\sum_{n\leq
0}|R_{0,u}e^{iK_{1}n}+S_{0,u}e^{iK_{2}n}|^{2}}.  \label{t_norm}
\end{equation}
To address the reciprocity of the transmission, we then define the \textit{
rectification factor} as
\begin{equation}
f=\frac{t(K_{1},K_{2},A,B)-t_{\mathrm{flip}}(K_{1},K_{2},A,B)}{
t(K_{1},K_{2},A,B)+t_{\mathrm{flip}}(K_{1},K_{2},A,B)},  \label{rectf}
\end{equation}
where $A,B$ is $T_{1,u},T_{2,u}$ in the case when the input is computed as a
function of the output, or $A,B$ is $R_{0,u},S_{0,u}$ in the case when the
output is computed as a function of the input. The $t_{\mathrm{flip}}$
notation indicates that the solution was computed with the potentials $U,V$
flipped across the midpoint between the $n=1,2$ sites. This is equivalent to
a solution with both $K_{1},K_{2}$ negative. Figure \ref{fNLIN} shows a plot
of $f$ as a function of $T_{1,u},T_{2,u}$ to demonstrate non-reciprocity of
the nonlinear system. Similar to what has been previously observed for the
Hamiltonian nonlinear asymmetric chains in~Ref. \cite{Lepri2011,Lepri2013},
and for the single $\mathcal{PT}$-symmetric chain in~Ref. \cite
{D'Ambroise2012}, the transmission asymmetry is evident. It is worthy to
note that there appears a set of near-unity values of the output-wave
parameters, $T_{1,u}$ and $T_{2,u}$, for which this asymmetry is most
pronounced.

If $\varepsilon =0$, i.e., the skew part is absent in potential (\ref{Pot}),
then the $K_{1},K_{2}$ branches are decoupled in the following sense.
Setting $S_{0,u}=0$ and $R_{0,u}\neq 0$ gives solutions for the remaining
amplitudes such that $T_{1,u},R_{u}\neq 0$ and $T_{2,u},S_{u}=0$. In other
words, if the $K_{1}$ branch is present, while $K_{2}$ is absent in the
incident part of the wave, then only $K_{1}$ will be present in the
reflected and transmitted waves. The same is true for $K_{2}$ if it is
originally present while $K_{1}$ is not. For $\varepsilon =0$, the results
demonstrate the reciprocity in the linear case $(\lambda =0)$, so that $f$
defined as per Eq. (\ref{rectf}) is always zero, for $t$ defined either as
in Eq. (\ref{t_amp}) or as in Eq. (\ref{t_norm}).

If $\varepsilon \neq 0$, then the $K_{1},K_{2}$ branches are coupled, hence
setting $S_{0,u}=0$ and $R_{0,u}\neq 0$ gives solutions for the remaining
amplitudes such that $T_{1,u},R_{u},T_{2,u},S_{u}\neq 0$. In other words, if
the $K_{1}$ branch is present and $K_{2}$ is absent in the incident part of
the wave, then both $K_{1},K_{2}$ will appear in the reflected and
transmitted parts of the wave. The same is naturally also true if $K_{2}$ is
incident in the absence of $K_{1}$.

Thus far we have considered extended wave solutions in the form of Eq. (\ref
{ansatz}) with real $K_{1},K_{2}$ satisfying Eq. (\ref{omega}). Such
solutions also exist in the case when one of $K_{1},K_{2}$ is real and one
is complex. As follows from Eq. (\ref{omega}), this happens when one of the
expressions $\frac{1}{2}(q\mp \sqrt{\kappa ^{2}-\gamma ^{2}})$ is in the
interval $(-1,1)$, while the other one is not. If $K\equiv x+iy$ with real $
x $ and $y$ is the complex wavenumber (either $K_{1}$ or $K_{2}$), then we
can write $\cos K=\cos x\cdot \cosh y-i\sin x\cdot \sinh y$. From Eq. (\ref
{omega}) we also have that $\sqrt{\kappa ^{2}-\gamma ^{2}}=\cos (K_{1})-\cos
(K_{2})$ is real. From here it follows that $x$ is a multiple of $\pi $ so
that $\cos K$ is real, and $y=\pm \mathrm{\ Arcosh}(|\cos \left( K\right) |)$
. The positive value of $y$ allows for solutions in the form given by Eq. (
\ref{ansatz}) to stay finite as $n\rightarrow \pm \infty $ (i.e., the wave
with complex $K$ is a localized one). In this case, one must set, at 
$\infty $, the incident amplitude associated with $K$ equal to
zero ($S_{0,u}=0$ if $K=K_{2}$ or $R_{0,u}=0$ if $K=K_{1}$). Since both
branches are generated when one branch is represented by the incident wave, the complex-$K$
contribution will appear in the reflected and transmitted parts. Figure \ref
{profLINKcomplex} shows a sample profile of such a physically relevant
solution with complex $K_{1}$.

\subsection{Stability}

\begin{figure}[tbp]
\centerline{
\includegraphics[width=9.5cm,angle=0,clip]{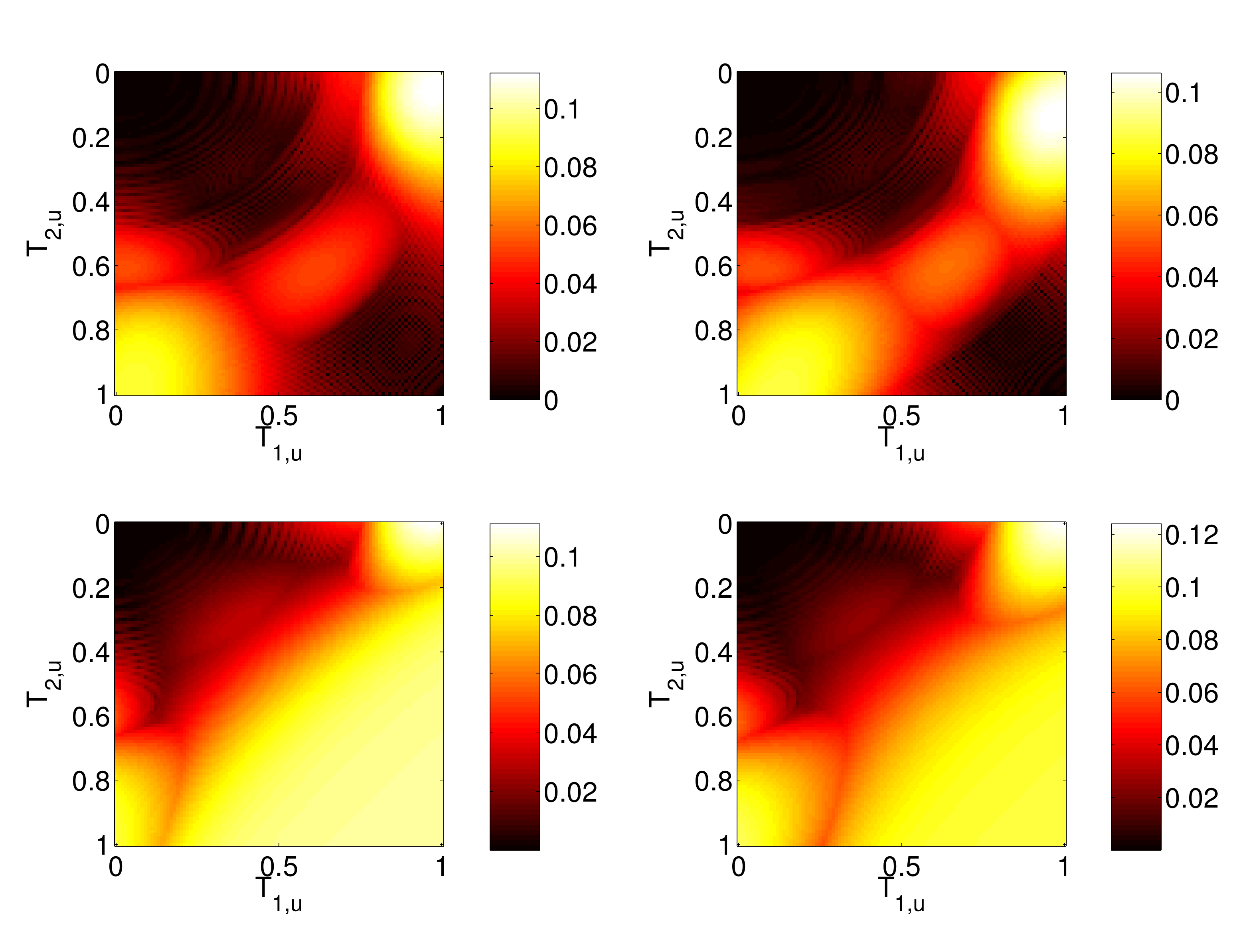}
}
\par
\caption{(Color online) Plots of $max(Re(i\nu))$ as a function of $T_{1,u}, T_{2,u}$ indicating the (in)-stability of the plane wave solutions that are computed as described in Section III.A.  Parameter values in each panel are: $\gamma=\varepsilon=0$ (top left), $\gamma=0$ and $\varepsilon=0.25$ (top right), $\gamma = 0.1$ and $\varepsilon = 0$ (bottom left), and $\gamma = \varepsilon = 0.1$.  All panels have values $K_1 = \pi/2, \lambda = 1, \kappa = 0.6, V_{0}=1$, and total ladder length of $200$.
\label{stab}}
\end{figure}

\begin{figure}[tbp]
\centerline{
\includegraphics[width=8cm,angle=0,clip]{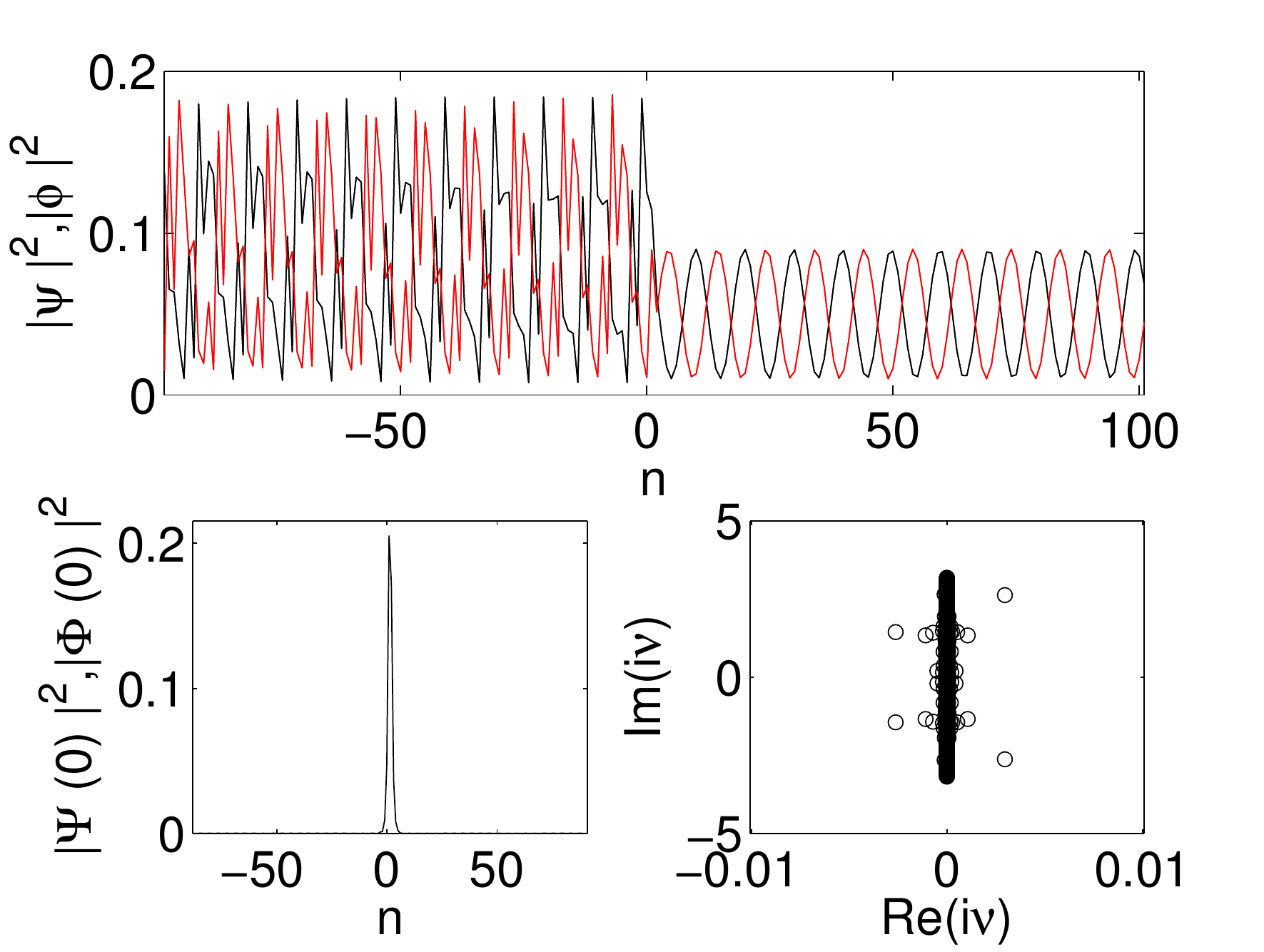}}
\centerline{
\includegraphics[width=8cm,angle=0,clip]{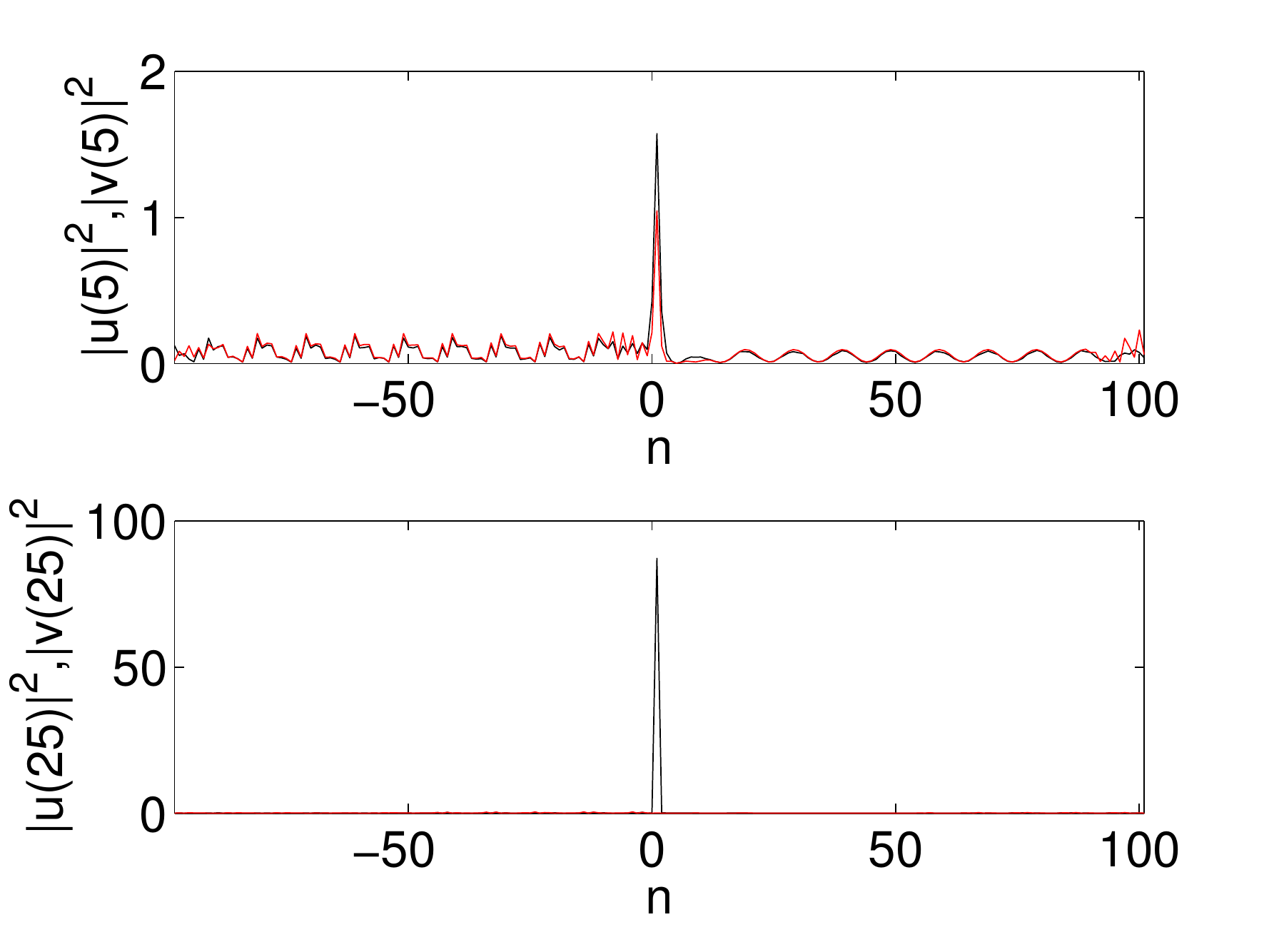}}
\par
\caption{(Color online) The top panel shows the stationary state $|\psi|^2, |\phi|^2$ (black, red) for parameter values $\kappa = 0.6, \gamma = 0.1, \varepsilon = 0.1, T_{1,u} = 0.1, T_{2,u} = 0.2$.  The solution is unstable and $i\nu$ is plotted in the complex plane in the right panel of the second row where $\nu$ are eigenvalues of $M$ in (\ref{M}).  An unstable eigenvector is plotted in the left panel of the second row.  The third and fourth rows show the evolution in time of the solution when perturbed in the direction of the eigenvector at $t=5$ and $t=25$.  For the wave function plots, blue corresponds to $u$ and green to $v$.
\label{unstabex}}
\end{figure}

To analyze the stability of the plane wave solutions we set $\{u_{n},v_{n}\}\equiv \{e^{iqz}(\phi _{n}+\delta \Phi_n(z)),e^{iqz}(\psi_{n}+ \delta\Psi_n(z))\}$ where $\phi_n, \psi_n$ are stationary solutions of (\ref{eq:stat}) in the form of (\ref{ansatz}) as described in Section III.A and $\delta>0$ is small.  Writing 
$\Psi _{n}=a_{n}e^{i\nu t}+b_{n}^*e^{-i\nu^{\ast}t}$ and 
$\Phi_{n}=c_{n}e^{i\nu t}+d_{n}^*e^{-i\nu ^{\ast}t}$ gives the linear system 
\begin{equation}
M\left[
\begin{array}{c}
a_{n} \\
b_{n} \\
c_{n} \\
d_{n}
\end{array}
\right] =\nu \left[
\begin{array}{c}
a_{n} \\
b_{n} \\
c_{n} \\
d_{n}
\end{array}
\right]
\end{equation}
with the nonzero entries of the $M$ matrix as follows
\begin{eqnarray}
M_{11} &=& diag(-q-i\gamma - U_n - 2\alpha_n |u_n|^2) + G \nonumber\\
M_{22} &=& diag(q-i\gamma + U_n + 2\alpha_n |u_n|^2) - G\nonumber\\
M_{33} &=& diag(-q+i\gamma-V_n-2\alpha_n |v_n|^2)+G\nonumber\\
M_{44} &=& diag(q+i\gamma+V_n+2\alpha_n|v_n|^2)-G\nonumber\\
M_{12} &=& -diag(\alpha_n u_n^2) = -M_{21}^*\nonumber\\
M_{34} &=& -diag(\alpha_nv_n^2) = -M_{43}^*\nonumber\\
M_{24} &=& M_{42} =  diag(\kappa) = -M_{13} = -M_{31} \label{M}
\end{eqnarray}
where $\alpha_n = \lambda \left( \delta _{n,1}+\delta _{n,2}\right)$ and $G$ is a sparse matrix with ones on the super- and sub-diagonals.  A stationary solution $\phi_n, \psi_n$ is then stable if $max(Re(i\nu)) > 0$.  Figure \ref{stab} shows the stability calculation as a function of the output amplitudes $T_{1,u}, T_{2,u}$ for stationary solutions according to Section III.A.  As expected, one sees a higher strength of instability for higher values of $\gamma, \varepsilon$. A typical example of an unstable plane wave solution is shown in Figure \ref{unstabex} with a corresponding eigenvalue/eigenvector pair and some snapshots of the propagation in time.  As time increases the amplitude of the unstable solution concentrates on one of the nonlinear nodes on the $u$-side of the ladder, i.e. on the gain side.  This behaviour is consistent with  previous results \cite{Ramezani2010, D'Ambroise2012}.

\section{Dynamical simulation of the wave-packet scattering}

We now turn to the scattering of finite-size wave packets on the central
core of the system, which is, obviously, another problem of physical
interest. We have performed simulations for the chains of finite lengths,
i.e., for $|n|\leq M$ (this means that each of the two chains is composed of
$2M+1$ sites). Open boundary conditions are enforced on both chains, namely $%
u_{-M-1}=u_{M+1}=0$ and $v_{-M-1}=v_{M+1}=0$. Initial conditions were taken
as a Gaussian wave packet, with the center placed at point $n_{0}<0$:
\begin{equation}
\left(
\begin{array}{c}
u_{n}(0) \\
v_{n}(0)
\end{array}
\right) =\left(
\begin{array}{c}
A \\
B
\end{array}
\right) \exp \left( -\frac{\left( n-n_{0}\right) ^{2}}{w}-iKn\right) ,
\label{initial}
\end{equation}
where width $w$ is large enough, with respect to the typical wavelength.
Initial amplitudes $A$ and $B$ are chosen according to Eq.~(\ref{eq: BA}).
The pulse created in the form of Eq. (\ref{initial}) will thus be traveling
with the group velocity determined by dispersion relation (\ref{omega}):
\begin{equation}
c_{\mathrm{gr}}=-\frac{dq}{dK}=2\sin K.  \label{gr}
\end{equation}
\

To understand the results, it is necessary to keep in mind that, as
matter of fact, we create a mixture of two modes that correspond, according
to dispersion relation (\ref{omega}), to the same $K$, with equal group
velocities (\ref{gr}). Such a compound pulse will be traveling as a whole,
featuring internal intra-chain oscillations at the spatial beating frequency
\begin{equation}
K_{\mathrm{beat}}=\pi /\sqrt{\kappa ^{2}-\gamma ^{2}},  \label{beat}
\end{equation}
as it follows from Eq. (\ref{omega}).

Here we report typical results of the simulations with initial condition (
\ref{initial}). To minimize the dispersive effects and, thus, the dependence
of the scattering on the initial position $n_{0}$, we focus here on the case
of $K=\pi /2$. Moreover, for given lattice size $M$, the simulation duration
$z_{\mathrm{fin}}$ is limited so as to avoid the hitting of 
the boundary sites by
the transmitted and reflected packets. Then, the wavepacket-transmission
coefficients for the two coupled chains, produced by the simulations in the
interval of $0<z<z_{\mathrm{fin}}$, are naturally defined as
\begin{eqnarray}
&&t_{u}=\frac{\sum_{n>n_{\ast }}|u_{n}(z_{\mathrm{fin}})|^{2}}{
\sum_{n<1}(|u_{n}(0)|^{2}+|v_{n}(0)|^{2})},  \label{tu} \\
&&t_{v}=\frac{\sum_{n>n_{\ast }}|v_{n}(z_{\mathrm{fin}})|^{2}}{
\sum_{n<1}(|u_{n}(0)|^{2}+|v_{n}(0)|^{2})},  \label{tv}
\end{eqnarray}
the total transmission being $t=t_{u}+t_{v}$. However, a problem
with this definition is that the power in the rightmost region may
grow, due to the contribution from the region near the central core,
where the waves may be trapped and amplified by the gain term,
because the nonlinearity may break its balance with the loss. To
avoid this, we measure the transmitted power far from the center in
a ``moving window" containing all the transmitted power, but
excluding the trapped fraction localized around $n=1,2
$. This is accomplished by extending the sums in Eqs.~(\ref{tu}) and (\ref
{tv}) to the region of $n>n_{\ast }$, with $n_{\ast }=c_{\ast }z+m$, where $
c_{\ast }$ is equal to or smaller than the group velocity $c_{\mathrm{gr}}$ (
$c_{\ast }=1.8$ is fixed henceforth), and $m$ is a suitable constant.

\begin{figure}[tbp]
\centerline{\includegraphics[width=0.55\textwidth,clip]{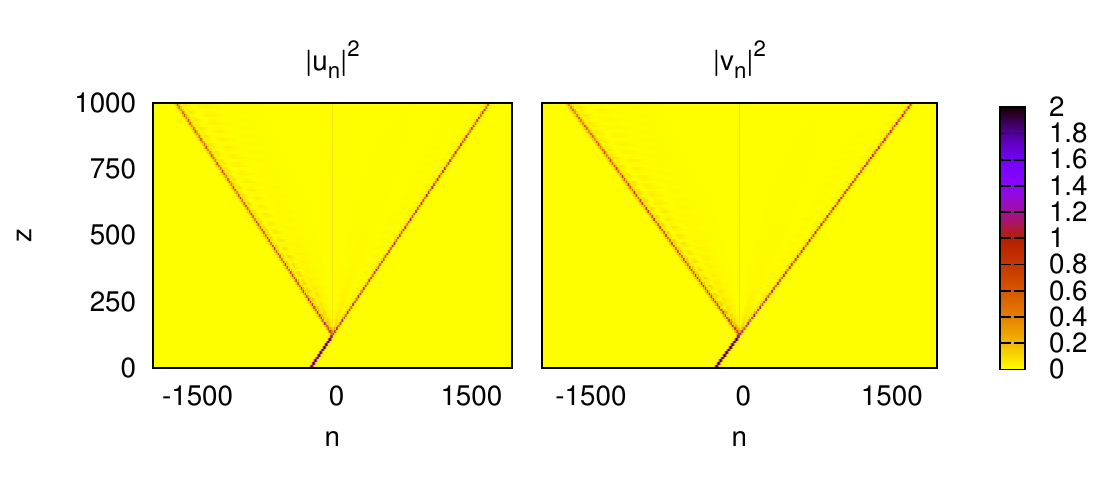}} 
\centerline{\includegraphics[width=0.55\textwidth,clip]{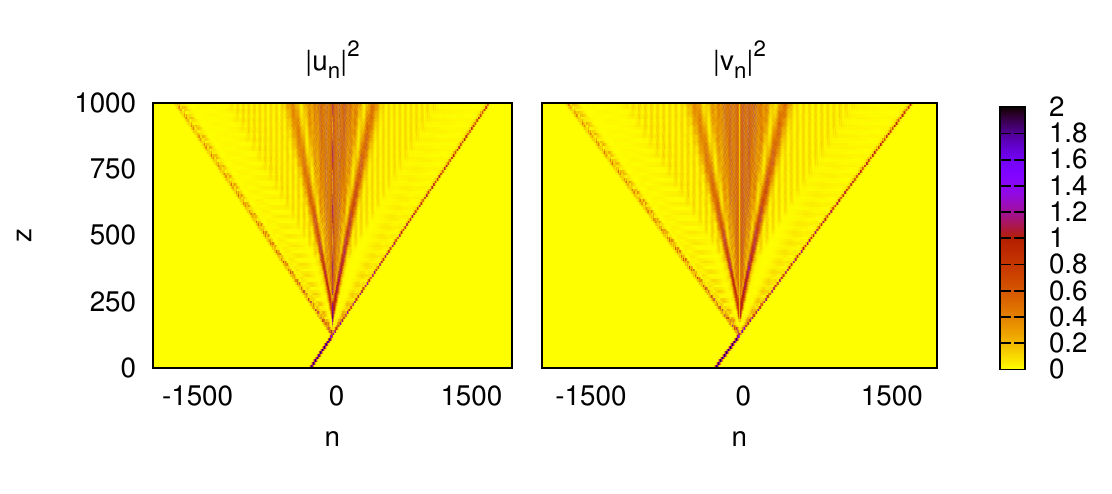}} 
\centerline{\includegraphics[width=0.55\textwidth,clip]{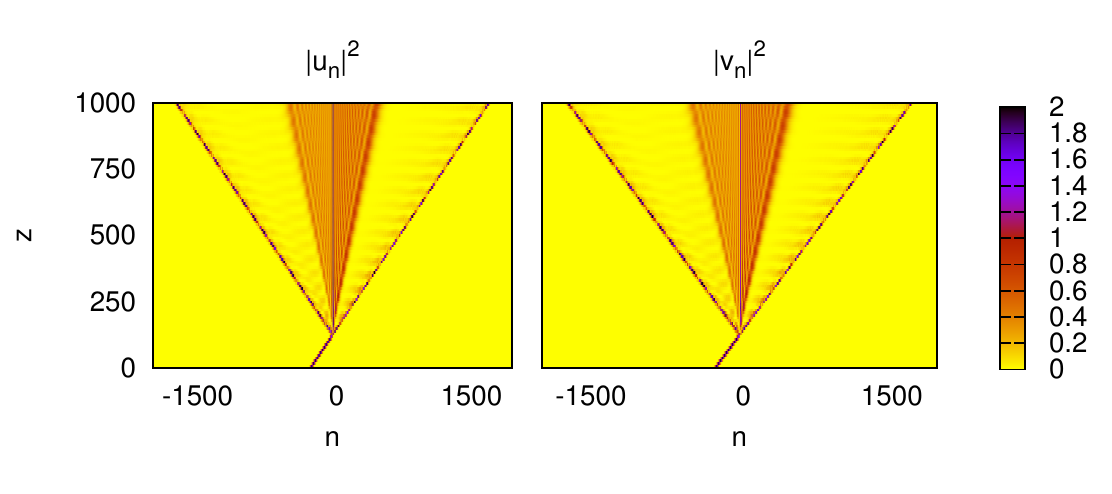}}
\caption{(Color online) Numerical simulations of the transmission of
Gaussian wavepackets for three values of the gain coefficient, $\protect
\gamma =0$, $0.05$ and $0.08$ (from top to bottom). Parameters are $\protect
\lambda =1$, $V_{0}=-2.5$, $\protect\kappa =0.1$, $K=\protect\pi /2$, $
\protect\varepsilon =0.05$, $M=2000$, $|A|^{2}=2$, $w=20$ and $n_{0}=-250$.}
\label{moderate}
\end{figure}

In Figs. \ref{moderate} and \ref{large}, we display results of the nonlinear system 
for different values of $\gamma $. In addition to the
transmission/reflection of the wavepacket, for $\gamma \neq 0$ we
also observe an ``after-effect" on the central sites of the ladder,
$n=1,2$, after the collision with the localized wavepacket.
The sites break the balance between the gain and loss, inducing the growth
of the power at the gain-carrying sites. The growth does not stay localized
at $n=1,2$, but rather expands to additional rungs of the ladder.

\begin{figure}[tbp]
\includegraphics[width=0.5\textwidth,clip]{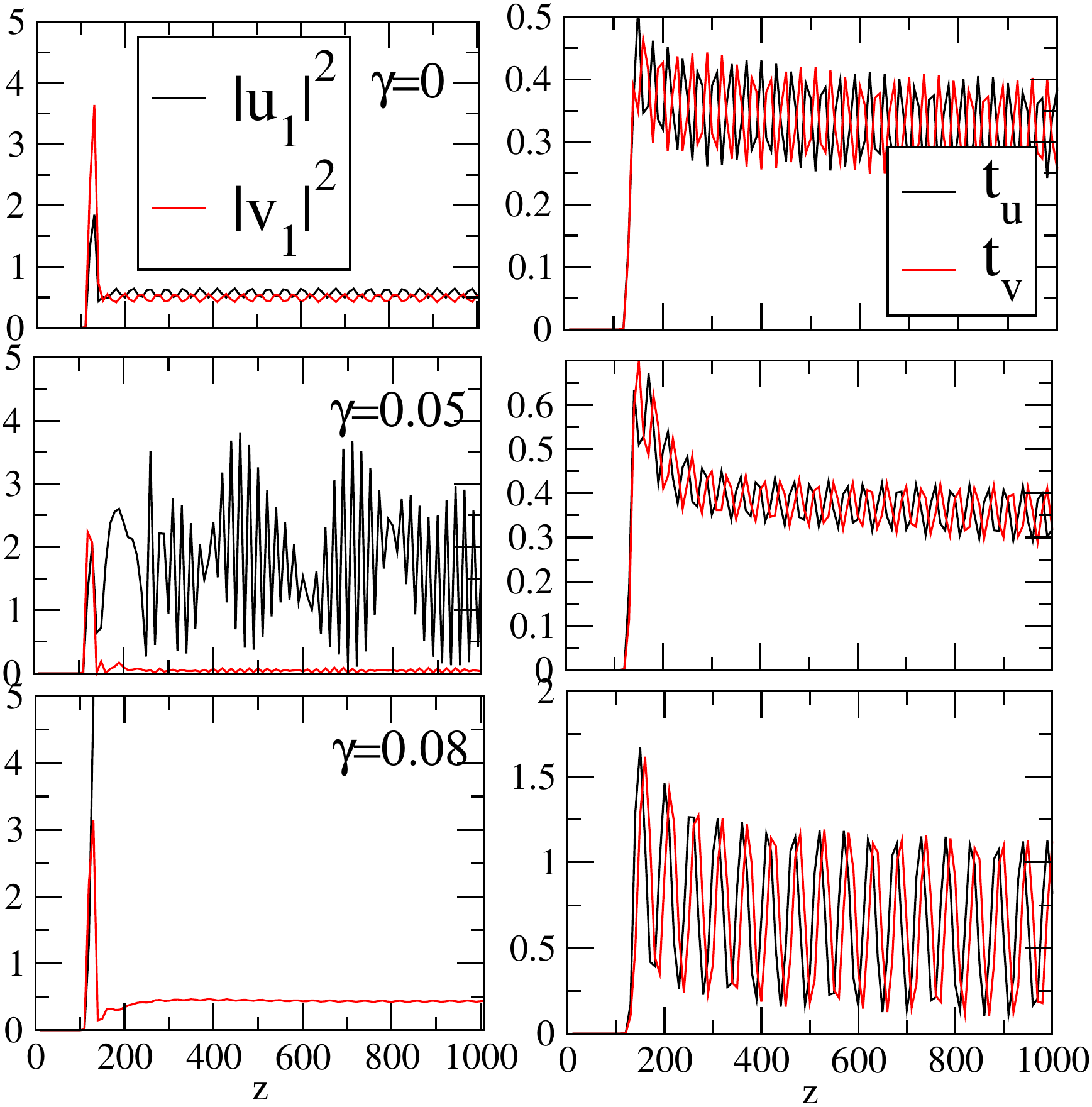}
\caption{(Color online) Simulations of the wave-packet scattering for
increasing gain strength $\protect\gamma $, with other parameters as in the
previous figure. Left panels: the evolution of the norm at the central site.
Right panels:\ the evolution of the transmitted intensities as defined by
Eqs.~(\protect\ref{tu}) and (\protect\ref{tv}). Note the steep
growth of $|u_{1}|^{2}$ observed in the left-bottom panel; outside the displayed window $|u_1|^2$ continues to grow.}
\label{central}
\end{figure}

\begin{figure}[tbp]
\includegraphics[width=0.45\textwidth,clip]{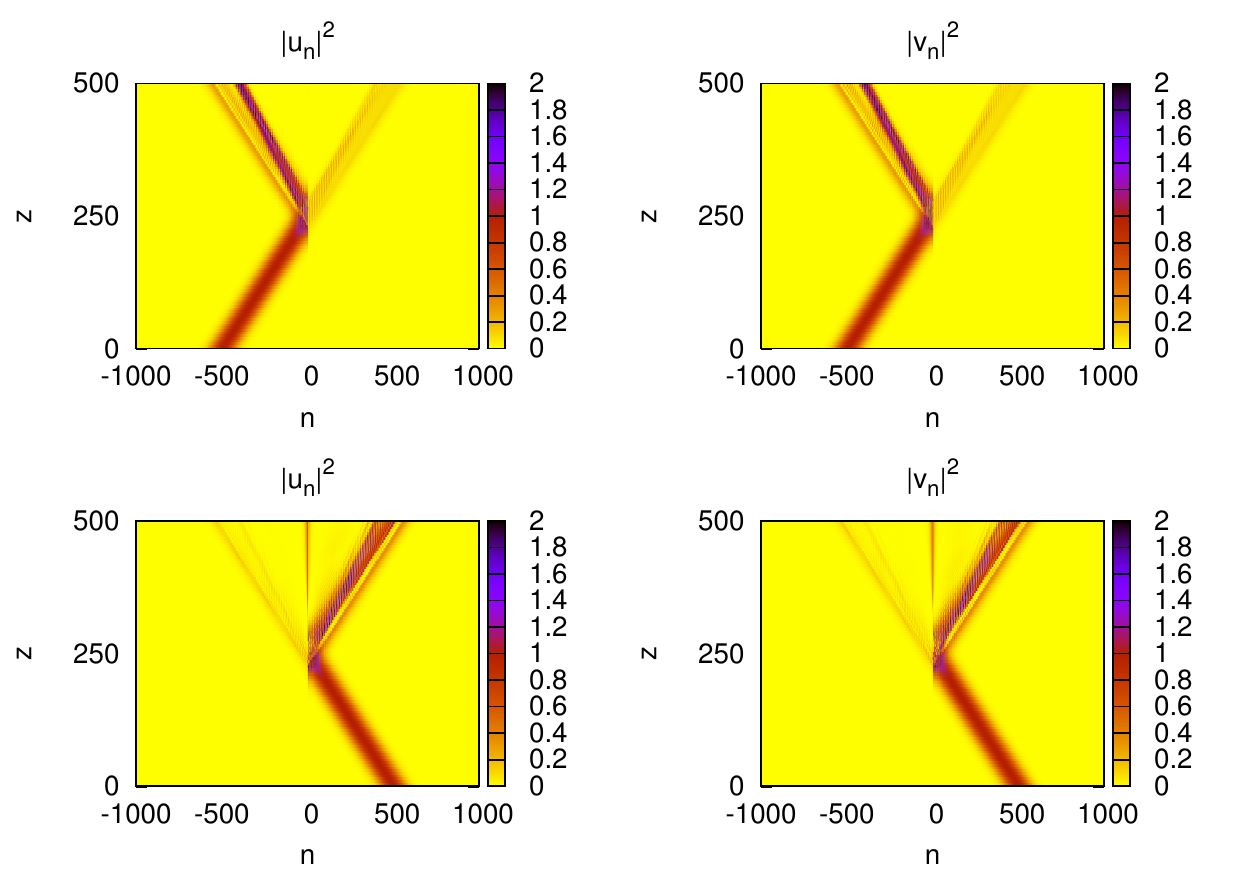} 
\includegraphics[width=0.45\textwidth,clip]{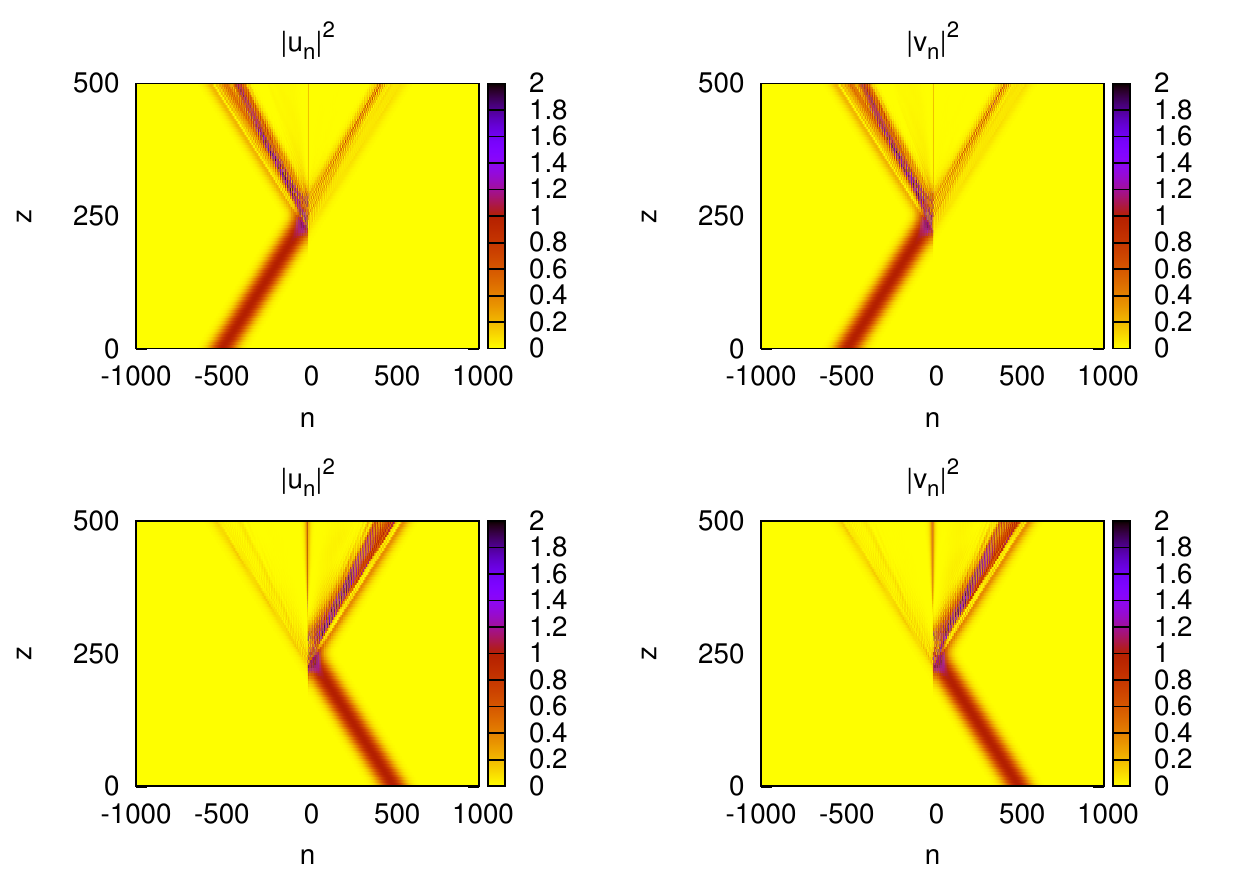}
\caption{(Color online) Spacetime plots of numerical simulations of the transmission of
Gaussian wavepackets for moderate and large nonlinearities, $\protect\lambda 
=0.75$ (top two rows) and $\protect\lambda =1.0$ (bottom two rows). Parameters are $\protect\kappa =0.6$, $\protect\gamma =0.1$, $
V_{0}=-2.5$, $K=\protect\pi /2$, $\protect\varepsilon =0.25$, $M=2000$, $
|A|^{2}=1$, $w=100$ and $n_{0}=-500$. }
\label{large}
\end{figure}

To monitor the effect of the interplay of the nonlinearity with the gain and
loss in the system, in the left panels of Fig.~\ref{central} we display the
evolution at the central sites. As $\gamma $ increases, the power
attains large values at these sites. The right panels show the
transmission as a function of evolution variable 
$t_{u,v}(z_{\mathrm{fin}})$, avoiding the
growing part, as described above. Note that these panels clearly demonstrate the expected oscillations at the
beating frequency given by Eq. (\ref{beat}).

\begin{figure}[tbp]
\includegraphics[width=0.45\textwidth,clip]{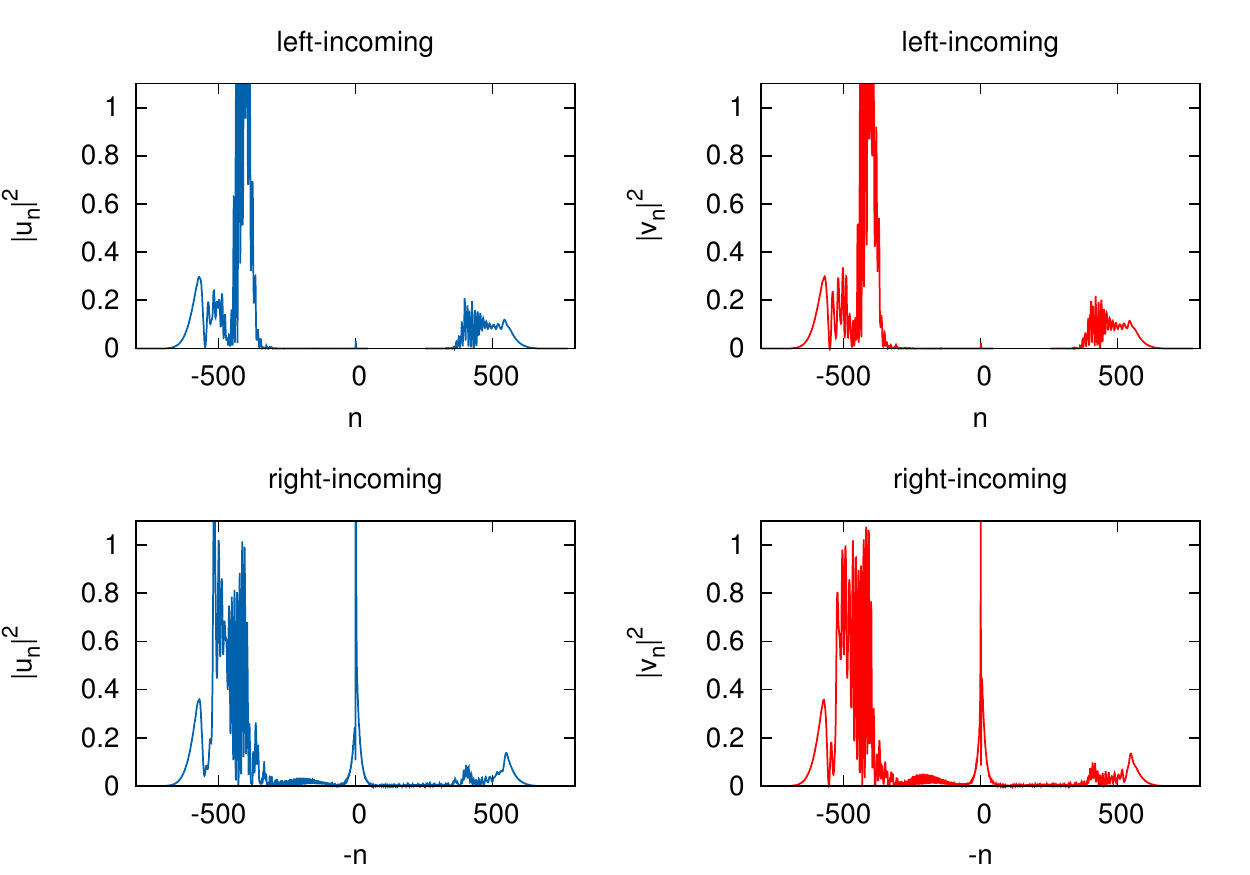} 
\includegraphics[width=0.45\textwidth,clip]{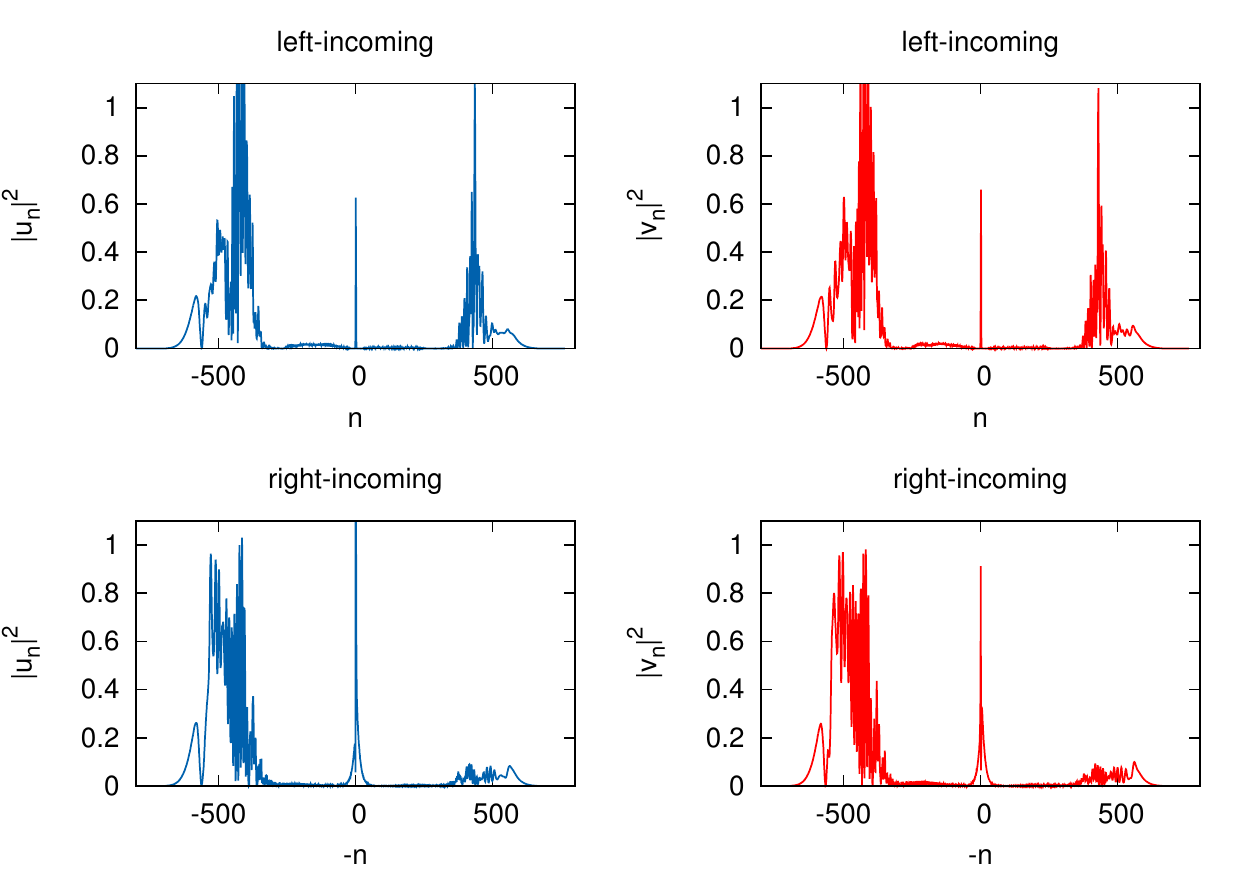}
\caption{(Color online) Profiles of the Gaussian wavepackets 
at a time after the packet has interacted with the nonlinear core.  
Parameter values are the same as those in Figure \ref{scatpt}.
Note that the horizontal axis
for the data referring to the right-incoming packets have been
mirror-reversed to facilitate the comparison with the left-incoming ones}
\label{scatpt}
\end{figure}

Next we address the issue of the asymmetric (non-reciprocal) transmission.
This is done by comparing the scattering for the same packet impinging on
the nonlinear core from the opposite direction. A noteworthy effect is seen
in that regard in Fig. \ref{scatpt}, where we compare increasing values of
the nonlinearity coefficient $\lambda $ (this is of course equivalent to
raising the input power). For moderate values ($\lambda =0.75$, in 
the left panels of Fig. \ref{scatpt}), reciprocity violations in the transmitted
intensities are manifest, as the outgoing pulses are different in their shape
and intensity. An important manifestation of the
non-reciprocity is that some energy remains trapped by the central segment only
for the right-incoming packets, but not for the left-incoming ones (which,
in turn, feature a stronger transmission). For a larger nonlinearity, $
\lambda =1.0$ (right panels in Fig. \ref{scatpt}), some power remains
trapped in both cases, although the amounts are different.
In this case too, the left-incoming packets feature a larger amount of the
transmission, while the right-incoming exhibit a considerably weaker
transmission and a larger trapping fraction. 

\section{Conclusions}

We have introduced and examined a ladder system with the $\mathcal{PT}$
-balanced combination of gain and loss uniformly distributed along
the pair of parallel chains, which are linearly coupled in the
transverse direction, and the core part, localized at two central
sites, which carry the linear potential and onsite nonlinearity. Two
branches of plane-wave solutions were found. The branches are
mixed at the central core if the potential functions are
skew-symmetric [$\varepsilon \neq 0$ in Eq. (\ref{Pot})], and they
stay uncoupled for $\varepsilon =0$. In the former case, asymmetric
transmission is observed, and is quantified by means of the
rectification factor, $f$, defined as per Eq. (\ref{rectf}). We have
also performed simulations of the interaction of incident
Gaussian wavepackets with the embedded core, similarly observing the
asymmetry of the transmission in the presence the nonlinearity at
the central sites. This asymmetry was quantified by suitable
transmissivities, and characteristic features of the evolution of
right- and left-incoming wavepackets were clarified through the
direct simulations.

As regards future work, it would be particularly relevant to explore
generalizations of\ the present settings to fully two-dimensional lattices,
a topic that has received relatively limited attention in the realm of $
\mathcal{PT}$-symmetric systems (see, e.g., Refs.~\cite{dark,midya}
for some recent examples). Another relevant possibility is to
consider, instead of the ``straight" ladder, with one chain carrying
the gain and the other -- the loss, an ``alternating" ladder, where
the gain-loss rungs would alternate with their loss-gain
counterparts , i.e., each gain node would be coupled to three
neighbors bearing the loss, and vice versa. Such settings are
currently under examination and will be presented in future
publications.

\end{document}